\documentclass[12pt]{article}  
\usepackage{epsfig} 
\usepackage{graphics} 
\usepackage{rotate}

\def \bg{\bigskip} 
\def \no{\noindent} 
\begin{document}
 \normalsize


  \bg

\centerline{ \large{\bf Equation of Motion of an Electric Charge }}  

\bg
{\centerline  {\bf Amos Harpaz$^{1,2}$  and Noam Soker$^2$ }} 

\bg

\no  1. Institute of Theoretical Physics, Technion, Haifa, ISRAEL  

\no 2. Department of Physics,  
Oranim, Tivon 36006, ISRAEL  

 \bg

    \no  phr89ah@tx.technion.ac.il 

\no soker@physics.technion.ac.il

\bg


\bg

\no {\bf   Abstract  }   

\bg

The appearance of the time derivative of the acceleration in the 
equation of motion (EOM) of an electric charge is studied.  It is 
shown that when an electric charge is accelerated, a stress force 
exists in the curved electric field of the accelerated charge, and this 
force is proportional to the acceleration.  This stress force acts as 
a reaction force which is responsible for the creation of the radiation 
(instead of the ``radiation reaction force"  that actually does not exist at 
low velocities).  Thus the initial acceleration should be supplied as 
an initial condition for the solution of the EOM of an electric charge.  
It is also shown that in certain cases, like periodic motions of an electric 
charge, the term that includes the time derivative of the acceleration, 
represents the stress reaction force.   

   \vfil

  \eject

\bg

\no {\bf 1. Introduction } 

\bg

The equation of motion (EOM) for a charged particle was calculated by 
Schott[1] in a three dimensional formalism, and by Dirac[2] in 
a four dimensional formalism:  
$$ma^{\mu} = F_{ext}^{\mu} + \Gamma^{\mu} = F_{ext}^{\mu} +{2 e^2\over
 3 c^3}( \dot{a}^{\mu} - {1\over c^2}v^{\mu} a_{\alpha}a^{\alpha})\ . \eqno(1) $$
The first term in $\Gamma$ (known as Schott term) includes the time 
derivative of the acceleration (a third time derivative of the position), 
and this raised the question as ``what is the role of the third time 
derivative in the EOM", as the presence of a third derivative  
 in the EOM demands the  definition of three initial conditions for 
the solution of such an equation, where in classical mechanics we 
usually need only two initial conditions for the solution of a regular EOM.  
Is the appearance of the third time derivative in the EOM just a sad accident 
that somehow should be bypassed, or is it a legitimate requirement that 
characterizes this kind of  motion?  We also notice that the second 
term in  $\Gamma$  equals the  power carried by the radiation (Larmor 
formula).   It is clear that Schott term should include the physical 
factors that create this power. 
 In certain physical situations, like a motion of an electric charge in a cyclotron, 
the third derivative of the position plays an important role.  It appears as a part of the 
``radiation force", ($F_{rad}$).   We return to this point in more details in section 3.  

One  way of bypassing the question of the third derivative is described 
in [3].  According to this suggestion, Schott term is isolated 
somehow from $\Gamma$ and is adapted to the acceleration term in the 
left side of the EOM (eq. 1).  To solve this equation in its new form, a 
multiplication by an integration factor ($e^{-\tau /\tau_0}$)  is needed.  
This term, which includes   the proper  time in an exponential form, 
leads later to divergent solutions that must be discarded.  It also causes an 
acceleartion that begins some time before the force that creates the 
acceleration acts, and this ``preacceleration" should  be discarded by 
causality considerations.   We shall discuss in detail the question of the 
third derivative  in section 2. 

Landau and Lifshitz[4, p. 203] also obtain the third derivative using a different 
approach.   They show that when a charge is accelerated by an external 
electric field, in addition to the force acting according to Lorentz formula, 
it is acted upon by a force $f = {2 e^2 \dot{a} \over 3 c^3}$, and by
multiplying this force by the velocity, and taking the average over time, they find 
the power, $P$, created by this force:  $P = {2 e^2 a^2 \over 3 c^3}$.   They call this 
force ``radiation damping", and argue that it represents the reaction of the radiation 
on the charge.  However, as was shown in [5], at low velocities the radiation is emitted 
in a plane perpendicular to the direction of motion, and no counter momentum 
is imparted to the charge by the radiation, and no radiation reaction force (or a radiation 
damping force) exists.  The reaction force to the motion of the accelerated charge 
should be found elsewhere.   Later we show that the reaction force is actually the 
stress force that exists in the curved electric field of the accelerated charge.  
The work performed in overcoming {\it this} reaction force is the source of 
energy carried by the radiation.

Until recently, $\Gamma$ term (also known as Abraham four vector) was considered 
as representing the radiation reaction force[6,7], and the fact that in a hyperbolic 
motion $\Gamma$ vanishes raised many questions, known as the ``energy balance 
paradox"[8].  The paradox is, that the vanishing of  $\Gamma$, leaves no 
radiation reaction force and no source for a work that may create the energy carried by 
the radiation.  

Following ref. [9] we adopt the approach given in a recently published work of Rohrlich[10]. 
 The equation of motion given in [10] is: 
$$ m_0\dot{v}^{\mu} = F^{\mu}_{ext} + F^{\mu}_{self}\  \ ,       \eqno(2)  $$ 
\no where $m_0$ is the physical rest mass involved in the process ($m_0 = m+\delta m$, 
 [10]), and  $F^{\mu}_{self}$  actually equals  Abraham 4-vector. 
$\delta m$  represents the electromagnetic  mass.   Rohrlich concludes that the fourth 
component of Schott term includes the power created by the reaction force,   
 and the vanishing of $\Gamma$ in a hyperbolic motion  shows that Schott 
term is the source of the power carried by the radiation.  However, the spatial part of 
Schott term (the force that creates this power) is not found, as the radiation reaction 
force does not exist.   In section 3 we show that there is a reaction force, but it is 
not  a {\it radiation} reaction force, but it is the stress force that exists in the curved electric 
field of the accelerated charge.  This reaction force should be included in the spatial part 
of Schott term, and the work done in overcoming this force creates the energy carried 
by the radiation.  

\bg

\no {\bf 2. Initial conditions  }

\bg

The answer  to the question about the role of the derivative of the 
acceleration demands analyssis of  all the factors 
involved in the motion of a charged particle.  It comes out that an 
important factor is usually overlooked - this factor is the stress 
force that exists in the curved electric field of an accelerated cahrged 
 particle[5].  The interaction of the curved electric field with the accelerated 
charge that induced the field appears as a reaction force that 
acts on the accelerated charge, and the work performed by the external 
force ($F_{ext}$) to overcome this reaction force is the source of the 
energy carried by the radiation[5].   The stress force is inversely 
proportional  to the radius of curvature, $R_c$, of the electric 
field (see eq. 9  further on).   For the simple case of a hyperbolic motion, 
 $R_c$ is given by:  
  $$R_c = {c^2\over a \sin \theta} \ \  , \eqno(3)   $$
\no  where  $a$ is the acceleration of the charged particle , and $\theta$ is 
the angle between the direction of motion and the direction of the field 
line at the location of the charge.  We find that the reaction force is 
proportional to the acceleration 
and  this shows that the value of the acceleration at any moment
 has an important role in the determination of the motion parameters, 
as it determines part of the forces that affect the motion.  Hence 
the initial value of the acceleration is needed to define the 
parameters of the motion, and the  appearance of the derivative of   the 
acceleration  in the EOM is not a sad mishap, but a legitimate 
requirement, needed for a complete solution of the physical 
situation. Later it is found[9], that when this stress force is 
calculated, its value should be inserted into the spatial part of Schott 
term, and then,  the power invested in overcoming this  force, balances 
the radiation power that appears in the second part of  $\Gamma^{\mu}$ 
that appears in eq. 1.  Thus $\Gamma^{\mu}$ is fully balanced as required 
(see [10]), and the ``energy balance paradox" is resolved.    Dirac treats 
the first term in eq. 1 ($ma^{\mu}$)  as the ``source of the kinetic energy", while 
Schott term is treated by him as the ``source of the acceleration energy".  This may 
be justified by noticing that this term includes the power created in overcoming the 
stress force (the reaction force) which is proportional to (and created by)  the acceleration. 

The situation described here resembles the situation in which a charge 
is supported statically in a gravitational field.  It is found[5] 
that this situation, which seems static is actually not static, but it is a 
steady state situation - the electric field of the charge is detached 
from the charge[11], and it falls in a free fall in the gravitational field. 
 The electric field 
becomes curved, and a stress force exists in this field. 
When the physical parameters are calculated for such a 
situation, for the most simple case (an electric charge located 
in a homogenous gravitational field[12]), it is found that in the 
close vicinity of the charge, the radius of curvature of the electric 
field,  ($R_c$),  equals (for the first approximation): 
$$R_c \simeq {c^2\over g \sin \theta}, \eqno(4)$$
\no where $g$ is the gravitational acceleration, and $\theta$ is the 
angle between the direction of $g$ and  the electric field line at the charge 
location.   This equation is similar to eq. 3, and this fact tempts us to 
deduce that there indeed exists an equivalence beween an acceleration 
of a charge in a free space, and a static location 
of a charge in a gravitational  field, and to the conclusion that a 
charge supported at rest in a gravitational field does radiate.  

\bg
\no {\bf 3. The Stress Force } 

\bg

Let us calculate in detail the power created by the stress force for the case of an hyperbolic 
motion. The equations for the electric field as calculated by Fulton and Rohrlich[6] using the 
retarded potentials method are given in cylindrical coordinates ($ z, \ \rho, \ \phi$), by: 
$$E_{\rho}={8e\alpha^2 \rho z\over \xi^3}              \eqno (5)$$  
$$E_z = {-4e \alpha^2\over \xi^3}[\alpha^2 +(ct)^2 + \rho^2 -z^2]                 \eqno (6)  $$ 
  $$B_{\phi} = {8e\alpha^2 \rho ct\over \xi^3}         \eqno (7)$$ 
 \no where $$ \xi^2 = [\alpha^2 + (ct)^2 - \rho^2 -z^2]^2 +(2\alpha \rho)^2 ,     \eqno (8)$$
\no and all other field components vanish. $z = \alpha=c^2/a$ is the location of the charge at time $t = 0$,  
and it is also the characteristic radius of curvature of the electric field.     The equation for  
the field lines was calculated by Singal[13], and they are drawn in  Fig. 1.  
 
\begin{figure}
\centering\epsfig{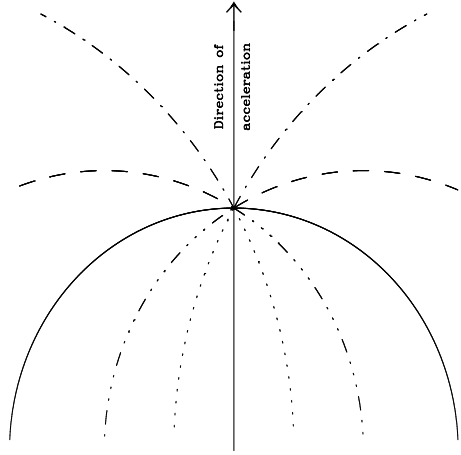}
\vskip 0.2 cm
\caption{\small The field lines of a charge moving with a constant acceleration}  
\end{figure} 

As can be expected, the field lines are curved and the radius of curvature, $R_c$, is inversely 
proportional to the acceleration and is given in eq. 3.   These equations were calculated earlier (in 
a three dimensional formalism) by Schott[1].  They were also calculated by Gupta and Padmanabhan[14]
who calculated first the electric field of the accelerated charge in its own system of reference, and then 
performed transformations to the free space system of reference.   They also show that by using the 
retarded variables ($x_{ret},\ t_{ret}$), they recover the equations usually given in the textsbooks[15, 16]. 

The  stress force density, $f_s$, is given (see [17] for more details) by: 
$$ f_s ={E^2\over 4\pi R_c}   \eqno(9) ,  $$ 
\no where  $E$ is the electric field.  When we substitute in eq. 9 for  $R_c$ 
 and for  $E = {e\over r^2}$ 
(which is correct for the close vicinity of the charge) we get for the stress force density: 
$$ f_s  = {e^2\over 4\pi}{a \sin \theta \over r^4 c^2}  \eqno(10). $$
 The component of this force in the direction of the acceleration is $-f_s \sin \delta$, where 
 $\delta$ is the angle between the field line and the direction of motion.  In the close vicinity of the charge, 
where the direction of the field lines did not change much, $\delta \simeq \theta$, and we find for the parallel 
component of the force density: 
$$ -f_s \sin \delta \simeq -f_s \sin \theta = {-e^2 a \sin^2\theta \over 4\pi c^2 r^4}   \eqno(11),   $$
 \no and we have to integrate this expression over a sphere around the charge. To avoid divergence 
at the center $(r \rightarrow 0)$, we take for the lower limit of the integration a small distance from the center,  
$c \Delta t$, where $\Delta t$ is infinitesimal.   As  the upper limit of the integration we take some large 
distance, $r_{up}$, where  $c\Delta t \ll r_{up} \ll c^2/a $.  The integration is carried in the free space system 
of reference, $S$,  which momentarily coincides with the system of reference of the charge at time $t = 0$.
 This integration yields the stress force $F_s$, by which the curved electric field acts on the accelerated charge.  
It acts against   the acceleration, and the external force that accelerates the charge, should overcome this stress
 force, in addition to the work it does in creating the kinetic energy of the charged particle.   This work is the 
source of the energy carried by the radiation.
\begin{eqnarray*}   
F_s = \int -f_s \sin \theta dV = {-e^2 a\over 4\pi c^2} \int_0^{2\pi} d \phi \int_0^{\pi} \sin^3 \theta d\theta 
\int_{c\Delta t}^{r_{up}} {dr\over r^2} = \cr  
  \left[{-2e^2 a\over 3 c^2r}\right]_{c\Delta t}^{r_{up}} = 
{-2 e^2 a\over 3 c^3 \Delta t}\left[1 - {c\Delta t\over r_{up}}\right]\ . \ \ \ \ \ \ \ \ \ \ \ (12) 
\end{eqnarray*}
  
Certainly, the second term in the square parenthessis can be ignored.  In order to find the power,
 $P_s$, created by this force, we multiply  $-F_s$ by the velocity of the charge at the time  $t = \Delta t$, 
$v = a \Delta t$, and find: 
$$ P_s = -F_s a \Delta t = {2e^2 a^2 \over 3 c^3}  \eqno(13)  $$ 
\no which is the power carried by the radiation (Larmor formula). This power should be 
inserted in the fourth component of Schott term, 
where it balances the power carried by the radiation. 

\bg
\no {\bf 4. A charge in a periodic motion } 

\bg

We present here two examples in which we find that $F_s$, which is responsible for 
the creation of the radiation, is proportional to $\dot{a}$.   

 Let us study   an oscillatory motion of a  
charge in a linear antenna whose length is $2D$ in the $x$ direction.  
The equations of motion for this case are:
$$ x = D \sin \omega t    \ \ ; \ \  
 v = \omega D \cos \omega t   \eqno(14.a)  $$  
$$ a = -\omega^2 D \sin \omega t  = -\omega^2 x  \ \ ; \ \  
 \dot{a} = -\omega^3 D \cos \omega t  =  -\omega^2 v \ .  \eqno(14.b)  $$ 

At time $t=\Delta t$, (where $\Delta t$ is infinitesimal), we use the 
 approximations (ignoring second and higher powers of $\Delta t$ and assumming 
 $ D\omega \ll c$):
$\sin \omega t \simeq \omega \Delta t$,  and $ \cos \omega t \simeq 1$, and find: 
$$ x = D \omega \Delta t \ \ ; \ \  
 v = D \omega   \eqno(15.a)  $$
$$ a = -D \omega^3  \Delta t \ \ ; \ \  
 \dot{a} = -D \omega^3  \ . \eqno(15.b)  $$
We find:  $ a = \dot{a} \Delta t $. (Note that in section 3 we had $a\rightarrow const., 
v \rightarrow 0$, while here, $a \rightarrow 0, v \rightarrow const.$   
The motion is linear, and we recall from eq. 12 that the reaction force $F_{reac}$ created in 
a linear acceleration motion at low velocity after short time $\Delta t$ is: 
$F_s = {-2e^2\over 3 c^3}{a\over \Delta t}$.  Using the relation we found above between 
$a$ and $\dot{a}$ we find $F_{reac} =   F_s = {-2e^2\over 3 c^3} \dot{a}$,  which is the 
expression used by Jackson [15, 17.8] for the radiation force $F_{rad}$, which is responsible 
for the creation of the radiation.  However, it was already shown that no radiation reaction force 
exists in a linear acceleration at low velocities, and $F_{reac}$, which should replace Jackson's 
$F_{rad}$, is the reaction force that exists as a stress force in 
 the curved electric field of 
the accelerated charge.  The force $-F_s$,    
should be inserted as the spatial part (three vector component) of  Schott term, 
and a term  $-F_s$, should be added to the accelerating force $F_{ext}$, 
which is needed to overcome the stress force. 

 As a second example,   
  consider the case in which a charge is moving in a circular motion like in a cyclotron.  
When a charge moving with a uniform velocity in  a free space enters into a homogenous magnetic field, 
$B$, which is aligned in a right angle to its direction of motion, it will be acted upon by a force  $F_B$, 
which is perpendicular both to its direction of motion and to the direction of $B$.  Since the force is 
always perpendicular to the charge's velocity, no work is performed by  $F_B$, and the charge will 
perform a circular motion due to the radial acceleration imparted to it.  If in such a motion 
no loss of energy takes place, this motion may continue forever.   However, due to the  
acceleration imparted to the charge, it radiates, and loses  energy through this radiation. 
The central force cannot perform  the work that creates the radiation, because it is 
always perpendicular to the direction of motion.     
In the equipment (cyclotron) in which this motion takes place, a force $F_{ext}$,  that acts 
along the direction of motion (along the circle) should act, where the work performed by  $F_{ext}$,  
creates the energy carried away by the radiation.  Usually, this force is implemented by an 
oscillatory electric field, whose frequency equals the frequency of the cyclotron. 
For simplicity, we assume that this force acts continueousely along the trajectory of the charge. 
It is clear that some force exists in the system which ``tries" to slow down the charge in its 
motion (it acts in an opposite direction to the velocity), and this force acts as a reaction force, 
$F_{reac}$, to the motion.   $F_{ext}$  should overcome  $F_{reac}$, and the question is:  what 
 is the source of  $F_{reac}$. 

The usually given answer to this question  is that $F_{reac}$ is a radiation reaction force, 
and it is called by Jackson[15, eq. 17.8],  $F_{rad}$.   
However, close inspection of the equation that gives the angular distribution of the radiation 
emitted from a charge moving  in a circular motion  [15,  eq. 14.44], shows that for low 
velocities of the charge in its circular motion ($\beta \ll 1$), the radiation is emitted in a 
plane perpendicular to the direction of the acceleration, and due to the symmetry of the 
radiation in this plane, no counter momentum is imparted by the radiation to the charge, 
and no radiation reaction force exists.  

\begin{figure}
\centering\epsfig{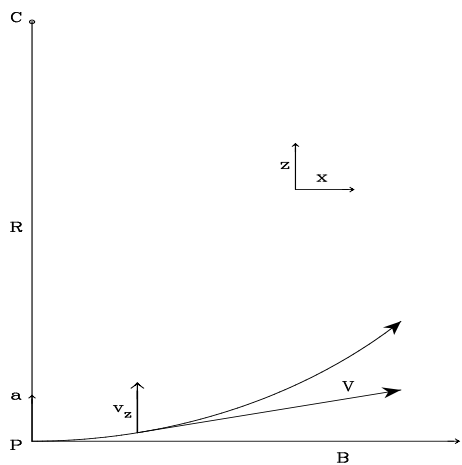}
\vskip 0.2 cm
\caption{\small A charge in cyclotron}  
\end{figure}

It was shown in [5], that when a charge is accelerated, its electric field becomes curved, 
and a stress force, $F_s$, exists in this curved field.  $F_s$  actually acts as a reaction 
force, which $F_{ext}$ has to overcome, and doing this,  $F_{ext}$ performs the work that 
creates the energy carried away by the radiation.  We are interested here in the component 
of $F_s$ in the counter direction of the velocity.  As shown in [5], the calculations are carried during 
an arbitrary infinitesimaly time interval, $\Delta t$.  In Fig. 2 we present the calculations:  The 
figure is drawn in the $x,\ z$  plane, as shown in the figure.  In this figure, the center of the 
circular orbit (with radius $R$), is the  point $C$.  At time $t =0$, the charge is located at the point  $P$. 
The velocity of the charge is $V$, and we analyze the situation in the system of reference 
$B$, which momentarily coincides with the system of reference of the charge at time: $t = 0$. 
 In this system the linear velocity of the charge vanishes, and a central velocity $V_z$ is 
created due to the central acceleration $a$.   At the end of the time interval $\Delta t$ a velocity 
$V_z = a \Delta t$ is created.  We recall from [5] (and from eq. 12 above)  that the component 
of the stress force parallel
to the acceleration is: $F_s = -{2e^2\over 3c^3}{a\over \Delta t}$.  We are interested in the 
component  $  F_{track}$ of $F_s$, which is  parallel (or anti-parallel) to the circular velocity 
$V$,  at time: $t = \Delta t$.     From trigonometric relations we find that at the time $t = \Delta t$:   
$${F_{track}\over F_s} = {V_z\over V} \ \ .  \eqno(16)   $$ 
The ratio of  $V_z$ to $V$ is: 
$$ {V_z\over V} = {a\over V} \Delta t = {-V\over R} \Delta t \ \ ,  \eqno(17)   $$ 
\no and we find for $F_{track}$:
$$F_{track} = F_s {V_z\over V} =-F_s {V\over R} \Delta t = { 2e^2\over 3c^3} a {V\over R} = 
{-2e^2 V^3\over 3c^3 R^2} \  \  , \eqno(18)  $$ 
\no where the last expression is obtained by substituting 
 $a =-\omega^2 R = { -V^2\over R}$.  
Using the same value for the acceleration, we have: 
$ -\omega^2 V ={-V^3\over R^2} = \dot{a}$, and we find that:   
$$F_{track} = { 2e^2 \over 3c^3 }\dot{a} \ \ ,  \eqno(19) $$ 
which equals the expression given by Jackson for $F_{rad}$,  the force that causes the 
creation of the radiation.   What we have found is that this force is not a radiation reaction, 
but it is a reaction force created by the stress force that exists in the curved electric field 
of the accelerated charge.  Multiplying $-F_{track}$  as given in eq. 16 by V, yield the power 
created by this force:
$$-F_{track} \cdot V = {2e^2 V^4\over 3 c^3 R^2} = {2e^2 a^2\over 3c^3} \  \  ,  \eqno(20) $$
\no which is the power carried by the radiation.  These calculations approve that also in 
the case of a circular motion (at low velocities), the stress force does the main role of 
the reaction force which is responsible for the creation of the radiation.  
 
\bg

\no {\bf 5.  A Charge Supported in a Gravitational Field  }  

\bg

The situation of a relative acceleration between a charge and its electric 
 field also exists for a charge supported statically in a gravitational 
field.   The electric field, which is not supported with the charge, falls 
in a free fall (see [11]), and there exists a relative acceleration between the charge 
and its field.   Prima facie, the situation seems static, but it is not.  
It is a steady state situation.  It is correct that in the lab system (the 
system of the supported charge), 
 the Poynting vector vanishes, but a Poynting  vector 
 is not an invariant, and cannot be taken as a covariant measure for 
the existence of radiation.  We demand that a covariant measure for the 
existence of radiation, is the non-vanishing of the Poynting vector in the 
system of reference defined by the geodesics.  Such a system is the one 
that falls freely parallel to the charge, and in this system, the 
supported charge is accelerated upward.  A magnetic field exists in this 
system, and the Poynting vector does not vanish (see Rohrlich[12]). 
 In this system, the electric field is curved, a stress force exists, and  calculations  
similar to those carried for the accelerated charge, yield the power carried by 
the radiation. The power can be calculated according to Larmor formula, 
where   we put $g$ for the acceleration, namely:  

$$ P = {2\over 3}{g^2 e^2\over c^3}  .  \eqno(13a)  $$

As a general measure for the radiation, we  take the 
four acceleration vector $ A^{\mu}$.  It is calculated  by: 

$$ A^{\mu} = D_4 U^{\mu} = {d\over dx^4} U^{\mu} + 
 \left\{{_4} {^{\mu}} {_{\nu}}\right\}U^{\nu} ,  \eqno(21)  $$

\no where $D$ represents a covariant drivative and
 $\left\{_{\gamma} {^{\mu}} {_{\nu}}\right\}$ is the Christoffel symbol.    
For a static charge $U^{\mu} =\left(\matrix { 0\cr c} \right)$.  
 The regular derivative in eq. 14 vanishes, and we are left with: 
  $$ A^{\mu} = \left\{{_4} {^{\mu}} {_4} \right\}U^4    \eqno(22)  $$
\no and for Larmor formula we should use: $ a^2 = g_{\mu \mu} A^{\mu}         
  A^{\mu}$, where  $a$ is the proper acceleration.   Let us examine  two cases:

(1) A homogenous static gravitational field. 

 (2) A Schwarzschild field.  

(1) Consider a  homegenous gravitational field directed downward in the 
$z (=x^3)$ direction, that generates an acceleration $g$. This field is 
characterized by  $u(z)$ (a function of $z$): 

 ${1\over u} = \cosh\sqrt{(1 - gz)^2 -1}$,  $u' = du/dz$, and  $g_{33} = (u'/g)^2$  (Rohrlich[12]).   
$$ A^3 = \left\{{_4} {^3} {_4}\right\}U^4 = g^2 {u\over u'} \eqno(23)  $$ 
\no and   
$$ g_{33}A^3A^3 = g^2 u^2 ,  \eqno(24) $$ 
\no which at  the charge location $(z \rightarrow 0 )$, yields   
 $g^2$ to be introduced in Larmor formula. 

(2) In a Schwarzschild field, the only relevant coordinate is $r (=x^1)$.     
We have: 
$$ A^1 = \left\{{_4} {^1} {_4}\right\} U^4 = {m\over r^2} (1 - {2m\over r}) . \eqno(25)  $$ 

The charge has an outward acceleration of $g (1 - 2m/r)$.  For Larmor formula 
we have to use   
 $a^2 = g_{11} A^1A^1 = g^2 (1 - 2m/r)$, and we find that 
for a static charge in a weak field,  Larmor formula yields a power very 
close to that radiated by an accelerated charge with an acceleration $g$. 

Let us calculate the electric field of a supported charge in a homogenous 
gravitational field.   Using equations given by Rohrlich[12], we  have 
(in cylindrical coordinates, $\rho, z, \phi$): 

$$ E_{\rho} = {8e \alpha^3 \rho u^2 \over \xi^3}  \eqno(26)  $$ 

$$ E_z ={-4e\alpha^3 uu' \over \xi^3}(\alpha^2(1 - u^2)+\rho^2) \eqno(27) $$ 

$$ \xi^2 = (\alpha^2(1-u^2) - \rho^2)^2 + 4\alpha^2\rho^2   , \eqno(28) $$  

\no where $u(z)$ and $u'$ are given above,  and here $\alpha = c^2/g$, 
which is the characteristic radius of curvature of the electric field 
close to the location of the charge.  
To calculate the fields in a horizontal plane passing through the location 
of the charge, we take $z = 0$, which gives: $u \rightarrow 1, \ \ 
 uu' \rightarrow 1/\alpha$.  These substitutions yields for the fields: 

$$ E_\rho = {e\over \rho^2}{1\over [1 + {\rho^2\over 4\alpha^2}]^{3/2}}  \eqno(26a) $$  
$$ E_z = {-e\over 2\alpha\rho}{1\over [1 + {\rho^2\over 4\alpha^2}]^{3/2}} 
 = {-eg\over 2\rho c^2 [1 + {\rho^2\over 4\alpha^2}]^{3/2}} .  \eqno(27a) $$  

If for the case described in section 3 (a charge moving in a hyperbolic motion) 
we 
calculate the electric fields of the charge in the plane  $z = \alpha$, at time $t = 0$, we find:
$$  E_{\rho} = {e\over \rho^2}{1\over [1 + {\rho^2 \over 4\alpha^2}]^{3/2}} ,    \eqno(5a)  $$ 
$$  E_z =   {-e\over 2\alpha \rho}{1\over [1 + {\rho^2 \over 4\alpha^2}]^{3/2}}  = 
   {-ea\over  2 \rho c^2}{1\over [1 + {\rho^2 \over 4\alpha^2}]^{3/2}}   ,   \eqno(6a)  $$
\no where here  $\alpha = c^2/a$, and we find that these equations are identical to 
 eqs.  26a, 27a,  except for the fact that $g$ is replaced by $a$.    

This means that if a charge accelerating in  hyperbolic motion  radiates, and the radiation 
in the plane perpendicular to the direction of motion is described by eqs. 5a, 6a, then 
the similar equations for the field of the charge supported in a homogenous gravitational 
field  (given by eqs. 26a, 27a), show that this charge also radiates. 

Usually, we consider the transverse field ($E_{trans}=E_z$) as describing the radiation field, 
as it falls in the close vicinity of the charge like 1/distance.  However, the presence of 
the extra term in the denominator  ($[1 + {\rho^2 \over 4\alpha^2}]^{3/2}$)  shows that 
for  large distances from the charge, the transverse field does 
not fall exactly like 1/distance,  as is usually found in the 
equations given in the textbooks[15,16].  The difference  of 
eqs. 5a, 6a, from the the equations given in the textbooks confused 
people who did not notice that equations  5a, 6a describe the 
electric field in the free space system of reference, while  
the equations given in the textbooks, with the nice neat dependence 
on 1/distance for the radiation, use the retarded coordinates 
($ x_{ret}, t_{ret}$).   (In ref [14], the authors show how the 
transformations between these two sets of coordinates work.)  
This problem confused Boulware[7], who tried to show that a 
coaccelerated observer will not observe  radiation from a 
hyperbolically accelerated charge.   By transforming from the retarded 
coordinates to the free space coordinates[7, eqs. III.18 - III.20]  Boulware  found that 
the nice dependence on 1/distance is lost, and from this fact he deduced  that the 
coaccelerated observer cannot observe the radiation.  This conclusion is correct for 
any observer, coaccelerated or not, and hence cannot be accepted  for the case 
of a coaccelerated observer. 

We find that the correct measure for the existence of radiation should be the ratio 
 $|{E_{trans}\over E_{long}}|  = |{E_{rad}\over E_{coul}}| > 1$, where $E_{trans}$  and 
$E_{long}$ are the transverse and longitudinal electric fields respectively, and $E_{rad}$  and 
 $E_{coul}$ ar the radiation field and the coulomb field respectively.  This ratio 
is found both in eqs. 5a, 6a, for a hyperbolically accelerated charge, and in 
eqs. 26a, 27a, for a charge statically supported in a homogenous 
gravitational field, at large distances from the charge. 

 What is the source of the energy carried by the  radiation in the gravitational field? 
 
  The charge is supported by a solid object, which is  static in  the GF 
(Gravitational Field).  This  solid objet is rigidly connected to the source of the
 GF.  Otherwise, it will fall in the GF, together with the ``supported"  charge.  
Actually,  the supporting object  is part of the object that creates the GF.

  The charge is static and no work  is done  on the charge.  However, the
 electric field of the charge is not static, and it falls in a  free fall in the GF. 
 With  no interaction between the electric field and the charge, the
      field would  follow a geodetic line and no work would 
       be needed to keep it  following the geodetic line. But the
     field is curved, and a stress force is implied.  The interaction
     between the curved field and the supported charge creates a
     force that contradicts the free fall.   In order to overcome this
force and cause the electric field to follow the geodetic lines, a 
 work is done on the electric field, and this work is  done  by the GF.  This work 
is the source of the energy  carried by the radiation.   The  energy carried away by 
the radiation is supplied by the  GF, that loses this energy.

\bg

\no {\bf 5.  Conclusions } 

\bg

By analyzing the forces that act on an accelerating charge we found that a stress force 
that exists in the curved electric field of the accelerated charge is proportional to the 
acceleration.   Hence, the initial value of the acceleration is needed as an initial condition 
for a complete solution of the charge's motion, and this property justifies the appearance 
of the third derivative of the position ($\dot{a}$) in the equation of motion of a charged 
particle.    The inclusion of the stress force as the spatial part of Schott term supplies the 
source of the energy carried by the radiation, and thus the ``energy balance paradox" 
is solved.  The stress force is calculated, and the work performed in overcoming it 
supplies the energy carried by the radiation. 
The similarity between the electric field of an accelerated charge, and that of a charge 
supported statically in a gravitational field, and the existence of relative acceleration 
between the charge and its electric field in both cases, leads to the conclusion that a 
 charge supported statically in a gravitational field radiates.  


\bg

\bg

\no {\bf {references:}}

\bg

\no 1. G. A. Schott, {\it Electromagnetic Radiation},  Cambridge University P., 1912. 

\no 2.  P.A.M. Dirac, {\it Proc. R. Society}, A167, 148,  1938.  


\no 3.  F. Rohrlich, {\it Classical Charged Particles}, Addison-Wesley,
 Reading, MA.,  1965.

\no 4.  L.D. Landau, E.M. Lifshitz, {\it Classical Theory of Fields}, Pergamon Press, 
3rd Ed., 1971.  

\no 5.  A. Harpaz, N. Soker, {\it Found. of Phys.}, 31, 935, 2001.

\no 6.  R. Fulton, F. Rohrlich, {\it  Ann. Phys.}, 9,  499,  1960.  

\no  7. D.G. Boulware, {\it  Ann. Phys.}, 124, 169, 1980. 

\no 8. C. Leibovitz, A. Peres, {\it  Ann. Phys.}, 25, 400, 1963.  

\no 9.  A. Harpaz, N. Soker, {\it Proc. Roy. Soc. London A}, submitted
(physics/0207038). 

\no 10. F. Rohrlich, {\it Amer. J. of Phys.}, 68, 1109, 2000.

 \no11. A. Harpaz, {\it Euro. J. of Phys.}, 23, 263, 2002. 

\no 12.  F. Rohrlich, {\it Annals of Phys.},   22, 169,  1963. 

\no 13.   A.K.  Singal, {\it Gen. Rel. Grav.}, 29, 1371, 1997. 

\no 14. A. Gupta, T. Padmanabhan, {\it Phys. Rev.}, D57, 7241, 1998. 

\no 15. J.D., Jackson, {\it Classical Electrodynamics}, 2nd Ed. (New York, Wiley), 1975. 

\no 16. A.K.H. Panofsky, M. Phillips, {\it Classical Electricity and Magnetism}, 2nd Ed. 
(Reading, MA., Addison-Wesley), 1964.

\no 17.  A. Harpaz, N. Soker, {\it Gen. Rel. Grav.}, 30, 1217, 1998.

  \end{document}